\documentclass[%
 reprint,
 amsmath,amssymb,
prb,
floatfix,
]{revtex4-1}

\usepackage{graphicx}
\usepackage{epstopdf}
\usepackage{dcolumn}
\usepackage{bm}
\usepackage{mathastext}
\usepackage{gensymb}
\setcitestyle{super}

\begin{document}
\setlength{\parskip}{0pt}

\preprint{APS/123-QED}

\title{High Pressure and Doping Studies on the Superconducting Antiperovskite $\bm{\mathrm{SrPt_3P}}$}

\author{BenMaan I. Jawdat$^{ }$}
\author{Bing Lv$^{ }$}
\author{Xiyu Zhu$^{ }$}
\author{Yuyi Xue$^{ }$}
\author{Ching-wu Chu$^{*}$}
\affiliation{
 $^{ }$ Texas Center for Superconductivity and Department of Physics, University of Houston, Houston, TX 77204-5002 \\
 $^{*}$ Lawrence Berkeley National Laboratory, 1 Cyclotron Road, Berkeley, CA 94720
}

\date{\today}

\begin{abstract}
We report the results of our investigation of $\mathrm{SrPt_3P}$, a recently discovered strong-coupling superconductor with $T_c$ = 8.4 K, by application of high physical pressure and by chemical doping. We study hole-doped $\mathrm{SrPt_3P}$, which was theoretically predicted to have a higher $T_c$, resistively, magnetically, and calorimetrically. Here we present the results of these studies and discuss their implications.
\end{abstract}

\pacs{Valid PACS appear here}
\maketitle



	In 2012, Takayama et al.\cite{Takayama2012} announced their discovery of $\mathrm{SrPt_3P}$, a centrosymmetric material that crystallizes in an antiperovskite structure similar to that of $\mathrm{CePt_3Si}$, with the notable difference that the distorted $\mathrm{Pt_6P}$ octahedral units are arranged antipolar rather than polar\cite{Takayama2012} which leads to centrosymmetry as opposed to non-centrosymmetry in $\mathrm{CePt_3Si}$\cite{Bauer2004}. This ternary platinum phosphide is a strong-coupling superconductor with a $T_c$ of $\sim$ 8.4 K\cite{Takayama2012}. Following the discovery of $\mathrm{SrPt_3P}$, several conflicting theoretical papers studying the electronic, vibrational, and thermodynamic properties of this material were published\cite{Subedi2013,Szczesniak2014, Kang2013,Chen2012}. One suggested that the structure might be tuneable by an appropriate choice of a substitute for phosphorus\cite{Subedi2013}; another suggested that the $T_c$ might increase with hole doping\cite{Nekrasov2012a}.
    
    Herein we report results of the chemical doping and high pressure effects on the properties of $\mathrm{SrPt_3P}$. In this study, the samples were prepared as follows. Stoichiometric amounts of  platinum powder, phosphorus powder, and strontium pieces were mixed together in a glove box under an argon atmosphere with total $O_2$ and moisture level less than 1 ppm. The mixture was pressed into a small pellet and loaded into a $Al_2O_3$ crucible. The crucible together with the pellet was then sealed in a clean quartz tube under vacuum. The whole assembly was then put inside a tube furnace for reaction. The reaction sequence is the same as reported by Takayama et al\cite{Takayama2012}. X-ray diffraction characterization of the material was carried out using a Panalytical X’pert diffractometer, magnetic measurements using the Quantum Design MPMS, and specific heat measurements using the Quantum Design PPMS. Resistivity measurements were made using a Linear Research LR-400 AC Resistance Bridge operated at 15.9 Hz and a modified probe in the MPMS and high pressure resistivity measurements were made using a BeCu piston-cylinder cell with a quasi-hydrostatic pressure medium, Fluorinert77.

\begin{figure}[b!]
  \centering
    \includegraphics[width=0.5\textwidth]{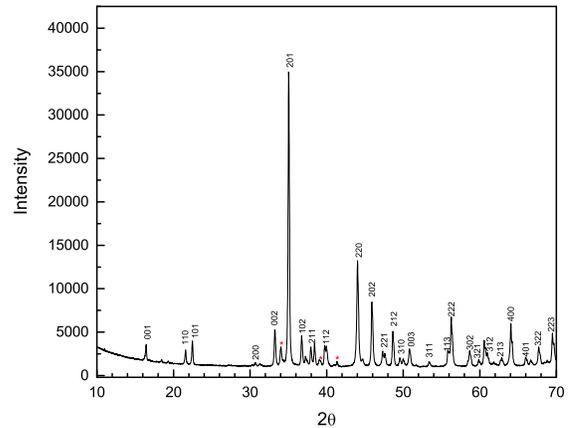}
  \caption{X-ray diffraction pattern for $SrPt_3P$. Unknown peaks are noted with a red mark.}
\end{figure}

The X-ray powder pattern of $\mathrm{SrPt_3P}$ is shown in Fig. 1. The sample is rather pure, as evidenced by the matching of almost all peaks to the structure proposed by Takayama et al. The relatively good quality of the sample was further verified by the rather sharp resistivity drop of the superconducting transition. The width of the resistive transition at 90\% of the drop is $\approx$ 0.31 K, and the diamagnetic shielding is $\approx$1.3 (without demagnetization corrections) as shown in Fig. 2 and its insets. The large difference between the field-cooled (FC) and zero-field-cooled (ZFC) data suggests type II superconductivity with possible strong field pinning.

We began our investigation of the properties of $\mathrm{SrPt_3P}$ by applying high physical pressure to the material.
\begin{figure}[h!]
  \centering
    \includegraphics[width=0.5\textwidth]{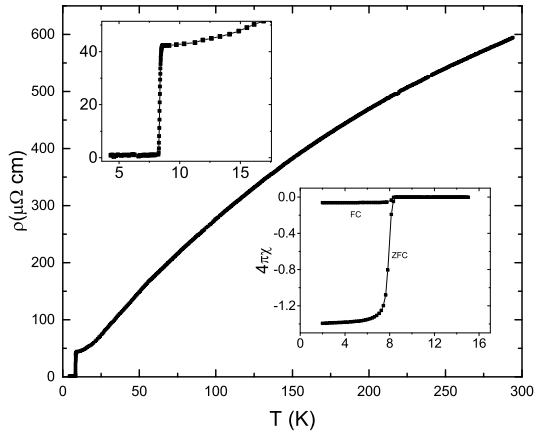}
  \caption{Temperature dependent resistivity and magnetic susceptibility of $SrPt_3P$. Inset (top left): the narrow resistive transition in an enlarged temperature scale from 0 to 20 K. Inset (bottom right): the magnetic susceptibility in ZFC and FC modes}
\end{figure}
\begin{figure}[h!]
  \centering
    \includegraphics[width=0.5\textwidth]{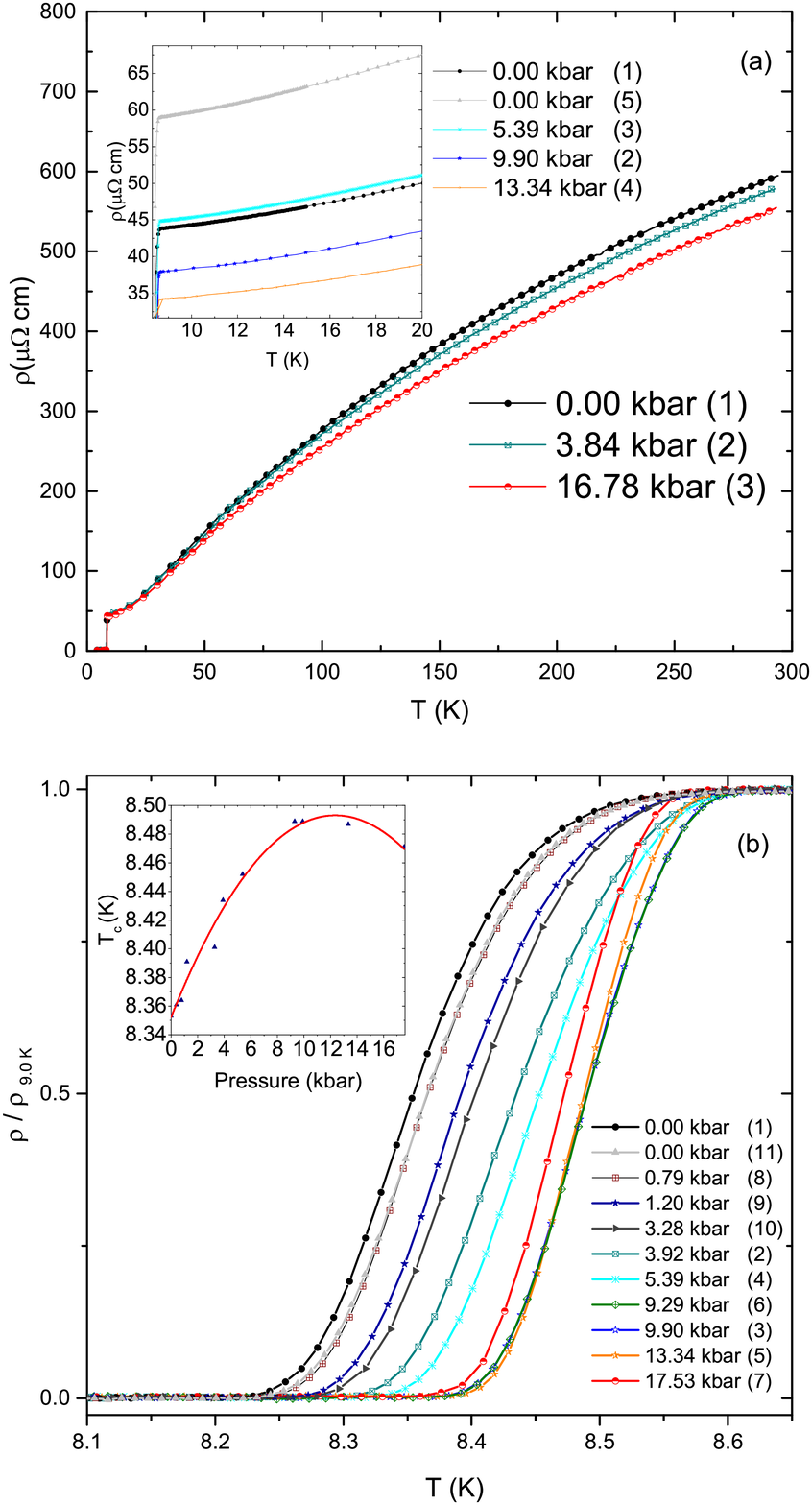}
  \caption{(a) Temperature dependent resistivity of $\mathrm{SrPt_3P}$ at different pressures. (b) Enlarged temperature scale of resistivity vs. temperature near $T_c$ for $SrPt_3P$ at different pressures (the sequential order of runs indicated by numbers in parenthesis). Insets: in (a) resistivity vs. temperature above $T_c$ for loading and unloading runs, and (b) $T_c$ vs pressure,}
\end{figure} 
The temperature dependent resistivity measurements under pressure are shown in Fig. 3. Taking the midpoint of the transition as the superconducting transition, we can clearly see an increase of the superconducting transition from 8.35 K at ambient pressure to a maximum 8.49 K at a pressure of 9.90 kbar (Fig. 3b). Further application of pressure beyond 9.90 kbar appears to lower the transition temperature, with the highest pressure of 17.53 kbar leading to a $T_c$ of 8.47 K. The change in slope of the superconducting transition at high pressure may be related to its width. Measurements were taken at several pressures upon unloading during the course of the experiment; we find that the trend is largely reversible  with a minor split ($<$0.01 K) between the pressure-increasing and pressure-decreasing branches.
\begin{figure}[t]
  \centering
    \includegraphics[width=0.5\textwidth]{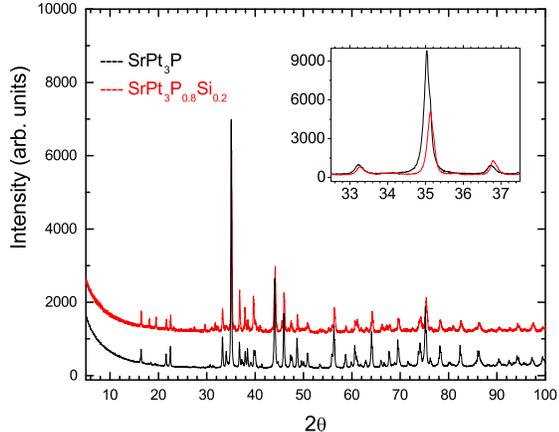}
  \caption{XRD results for the $SrPt_3P$ and $SrPt_3P_{0.8}Si_{0.2}$ samples. Inset: shift of the (201) XRD peak with increasing Si content.}
\end{figure} Upon unloading to ambient pressure, the $T_c$ does not recover its original value and instead is slightly higher. Irreversible defects may have been introduced to the sample during the course of applying pressure as evidenced by our observation that, upon unloading pressure, the resistivity above $T_c$ is higher than before loading (Fig. 3a, inset). Previous reports characterize $\mathrm{SrPt_3P}$ as a traditional electron-phonon superconductor in the strong-coupling regime\cite{Takayama2012,Subedi2013}; therefore, the increase in $T_c$ upon application of high pressure can be explained by an increase in the characteristic phonon energy $\Theta_D$\cite{Lorenz2005a}. The observed decrease in $T_c$ after the maximum is reached could be viewed as the result of a competition between an increasing Debye temperature and a decreasing density of states at the Fermi level\cite{McMillan1968, Allen1975}.
\begin{figure}[b]
  \centering
    \includegraphics[width=0.5\textwidth]{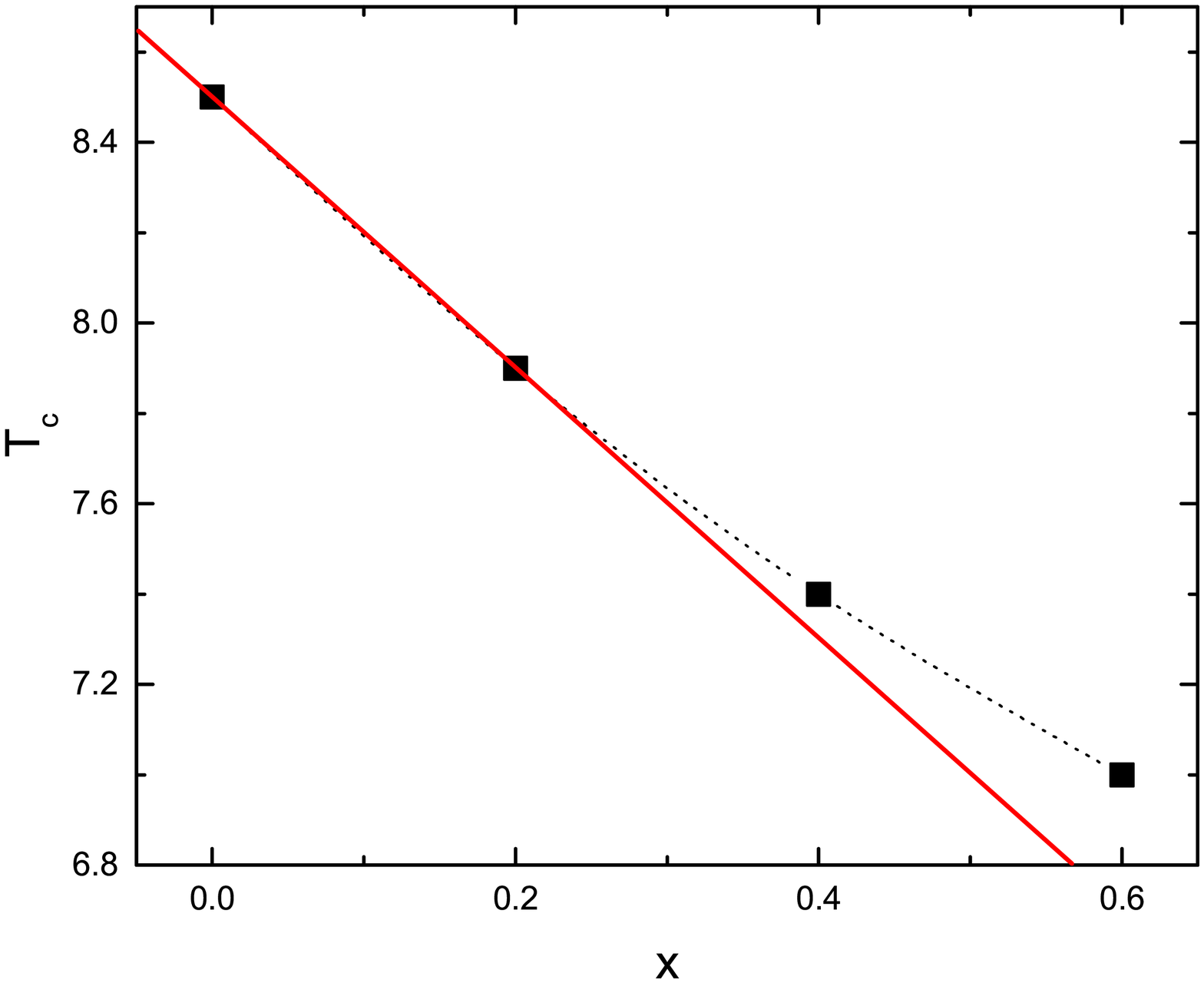}
  \caption{$T_c$ vs x for $\mathrm{SrPt_3P_{1-x}Si_x}$}
\end{figure}

Chemical doping was also carried out for the $\mathrm{SrPt_3P}$ compound. We selected Si as the dopant for two reasons: 1) the size of the Si is very close to that of P, minimizing any possible interference due to size difference with the formation of the material; 2) the doping of Si will likely introduce more hole carriers and will allow the study of hole-doping effects theoretically proposed by Nekrasov et al.\cite{Nekrasov2012a}. Fig. 4 shows the results of X-ray diffraction (XRD) with Si doping. The XRD spectrum of the un-doped $\mathrm{SrPt_3P}$ sample is compared with that of the Si-doped sample (in red). It is clear from this comparison that Si can replace P up to 20\% without inducing an impurity phase, as evidenced by the decrease of the lattice parameter show in the inset of Fig. 4. Higher Si-doping in $\mathrm{SrPt_3P_{1-x}Si_x}$ with x = 0.4 and 0.6 has also been carried out. Our XRD results clearly show a phase separation with the impurity phase increasing with x. This can also be seen in the deviation of $T_c$ from linearity with x (Fig. 5).

Temperature dependent magnetization and resistivity measurements were performed on samples with x = 0 and x = 0.2. Figure 6a shows that the onset transition temperature of the parent compound $\mathrm{SrPt_3P}$ is roughly 8.4 K. 

\begin{figure}[t!]
  \centering
   \centerline{\includegraphics[width=0.6\textwidth]{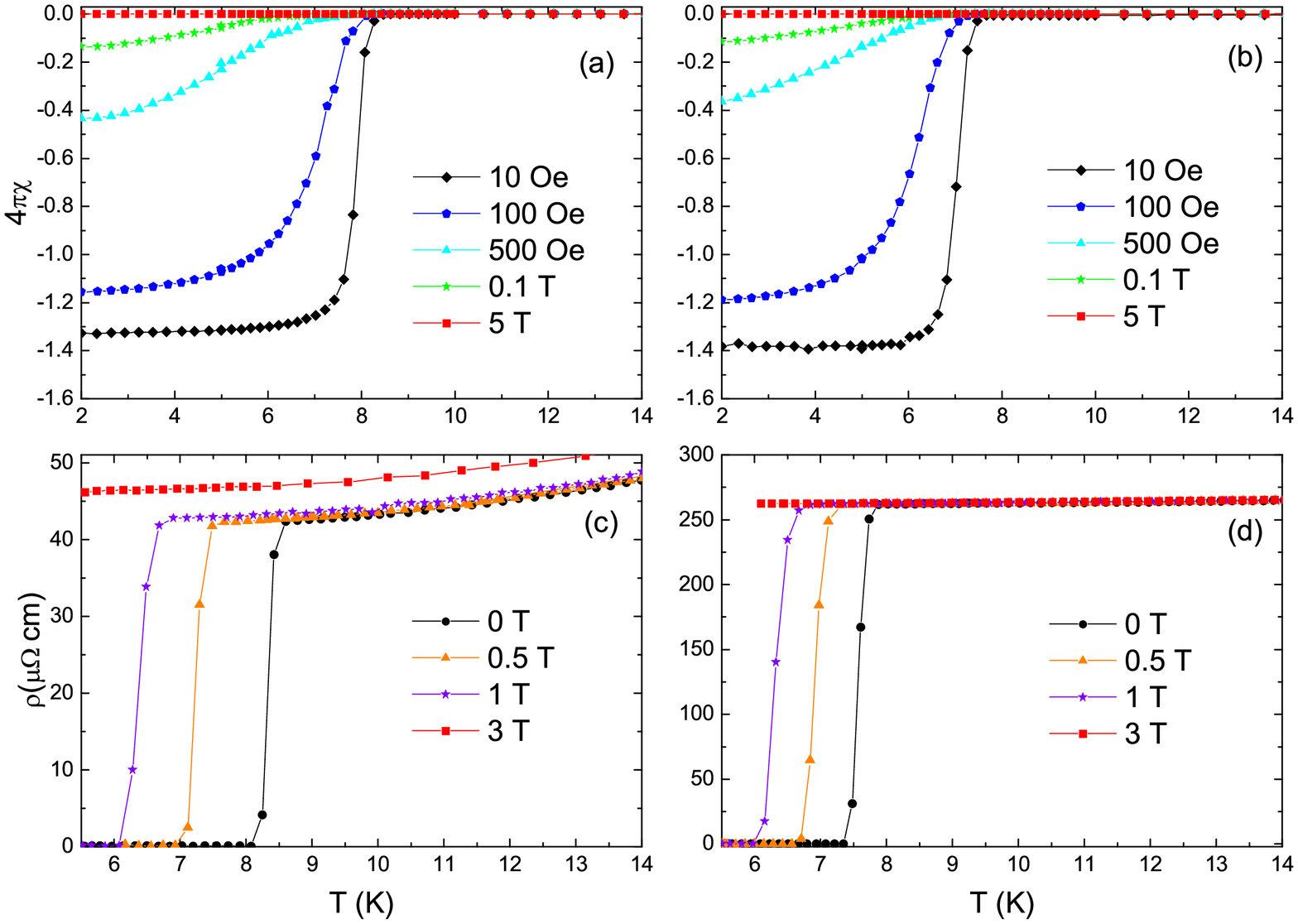}}
  \caption{Top: Magnetic susceptibility vs. temperature at different magnetic fields for (a) $\mathrm{SrPt_3P}$, (b) $SrPt_3P_{0.8}Si_{0.2}$; resistivity vs. temperature with different magnetic fields applied for (c) $SrPt_3P$, (d) $SrPt_3P_{0.8}Si_{0.2}$.}
\end{figure}
From our resistivity measurements, we observe suppression of superconductivity with a magnetic field of 3 T to below the lowest temperature measured. The temperature dependent magnetization of x = 0.2 is shown in Fig. 6b with an onset transition temperature of $\sim$7.8 K. We have also performed magnetization measurements on samples with x = 0.4 and x = 0.6, and observed that the $T_c$ decreases to 7.5 K and 7.0 K with increasing Si doping, respectively.

Bulk superconductivity was established in both the parent compound $\mathrm{SrPt_3P}$ and the sample with 20\% Si, $\mathrm{SrPt_3P_{0.8}Si_{0.2}}$, by measuring the specific heat of both the parent compound $\mathrm{SrPt_3P}$ and the sample with 20\% Si, $\mathrm{SrPt_3P_{0.8}Si_{0.2}}$. We apply the $\alpha$-model\cite{Padamsee1973, Johnston2013} to extract further information from the calorimetric data, which extends BCS theory by introducing a variable parameter $\alpha$ = $\Delta_0$/$k_B T_c$. Subtracting the normal state specific heat $\mathrm{C_N}$ (achieved by applying 7T magnetic field) from the superconducting state, we can arrive at a plot of $\Delta$C/T vs. T (Fig. 7). From this plot (knowing that the actual value of $\gamma = \Delta C / T - C_{el}(T)/T$ must lie between the lowest measured value of $\Delta C/T$ and the 0 K $\alpha$-model value) we can estimate the normal state $\gamma$ values for both samples. For the x = 0 sample, we find $\gamma$ = 6.31 $\pm$ 0.10 $mJ/mol\cdot K^2$, and for the x = 0.2 sample we find $\gamma$ = 7.77 $\pm$ 0.73 $mJ/mol\cdot K^2$. We arrive at $\alpha$ = 2.4 for x=0 and $\alpha$ = 2.2 for x=0.2. From this information we can extract $\Delta_0$ = 1.65 meV for $SrPt_3P$, which is agreeable with the report of $\Delta_0$ = 1.58 meV by Khasanov et al.\cite{Khasanov2014a}; we find $\Delta_0$ = 1.28 meV for $SrPt_3P_{0.8}Si_{0.2}$.
\begin{figure}[htp]
  \centering
    \includegraphics[width=0.38\textwidth]{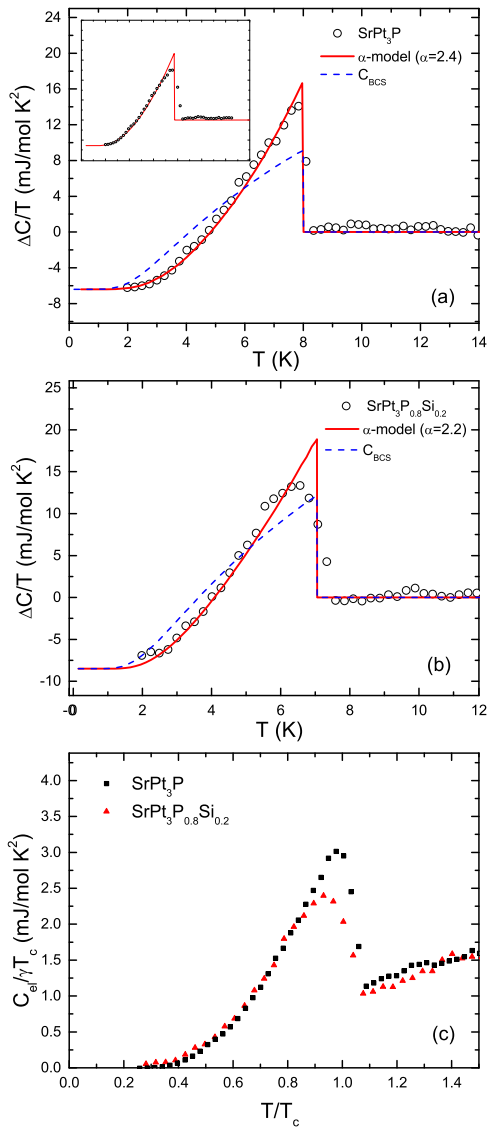}
  \caption{$\Delta$C/T vs. T for (a) $SrPt_3P$ and (b) $\mathrm{SrPt_3P_{0.8}Si_{0.2}}$. (c) $C_{el}/\gamma T_c$ vs. $T/T_c$ for both $SrPt_3P$ (broadened) and $\mathrm{SrPt_3P_{0.8}Si_{0.2}}$. The red lines represents the $\alpha$-model, and the blue lines represents the BCS-theory. Inset: $\Delta$C/T vs. T for $SrPt_3P$ with a broadened transition temperature width.}
\end{figure}
It can be seen from the specific heat data that the $T_c$ of the x=0.2 doped sample is broader than that of the undoped sample. For x = 0.2, the width is roughly 1.0 K, whereas for x = 0, the width is roughly 0.5 K. To verify that the difference in $\alpha$ value obtained from our fitting is not an artifact due to this difference between the samples, we artificially broadened the data of the x = 0 sample by assuming a 20\% content with 5\% lower $T_c$. We assume the same temperature dependence below $T_c$ for the lower $T_c$ component. This assumption results in a transition with a width that is the same as the x = 0.2 sample (1.0 K). This modified data is still well-fitted by $\alpha$ = 2.4 (Fig 8a. inset). Comparing the scaled specific heat $C_{el}/\gamma T_c$ vs. $T/T_c$ (Fig. 7c) for the x=0 data and the x=0.2 data with equal transition temperature width, we observe a significant decrease in the specific heat jump $\Delta C/ \gamma T_c$ at $T_c$ from $\approx$ 1.87 for the undoped sample to $\approx$ 1.36 for the x = 0.2 doped sample.

	The apparent increase of $N(\epsilon_F)$ with Si doping is in agreement with previous electronic structure calculations showing the Fermi level located on a negative slope in the density of states\cite{Nekrasov2012a}. The decrease in $T_c$ despite this increase in $N(\epsilon_F)$ supports the suggestion that the $T_c$ does not simply scale with $N(\epsilon_F)$\cite{Takayama2012}. The observed decrease in $\alpha$ could explain the observed decrease in $T_c$, and is in agreement with a previous report that decreasing coupling strength suppressed $T_c$ despite an increase in carrier density\cite{Kasahara2009}. 

	In conclusion, we investigated the superconductivity in $\mathrm{SrPt_3P}$ resistively, magnetically, and calorimetrically, including under the application of high physical pressure and with the partial replacement of phosphorus with silicon. The high pressure investigation yielded the result that the superconducting transition temperature increases by a maximum of approximately 0.14 K upon application of pressure up to 9.90 kbar, with an apparent decrease in $T_c$ beyond that pressure. Contrary to the previous predictions \cite{Nekrasov2012a}, hole doping with Si results in a systematic decrease of $T_c$, despite an apparent increase in $\gamma_N$. Our specific heat measurements demonstrate the bulk nature of the superconductivity in the parent compound as well as the silicon doped sample with x = 0.2; furthermore, we observe a decrease in $\alpha$ as well as $\Delta C / \gamma T_c$, implying that a decreasing coupling strength may be responsible for the suppression of $T_c$. We suggest that the observed decrease in coupling strength, combined with the suppression of $T_c$, indicates that the vibrational modes and hence the $T_c$ in this system may be highly sensitive to and tunable by modification of the details of the crystal structure.

\begin{acknowledgements}
The work in Houston is supported, in part, by U.S. Air Force Office of Scientific Research Grant FA9550-09-1-0656, the T.L.L. Temple Foundation, the John J. and Rebecca Moores Endowment, and the State of Texas through the Texas Center for Superconductivity at the University of Houston.

\end{acknowledgements}

\bibliography{library.bib}
\end{document}